\documentclass[journal,onecolumn]{IEEEtran}

\ifCLASSINFOpdf
\else
   \usepackage[dvips]{graphicx}
\fi
\usepackage{url}

\usepackage{supertabular}
\usepackage{graphicx}
\usepackage{amsmath,empheq}
\newtheorem{theorem}{Theorem}
\usepackage{algorithm}
\usepackage{algorithmicx}
\usepackage{algpseudocode}
\usepackage{booktabs}
\usepackage{mathtools}
\usepackage{amssymb}
\usepackage[table]{xcolor}
\mathtoolsset{showonlyrefs} 
\definecolor{light-gray}{HTML}{FFFFFF}
\definecolor{light-cyan}{HTML}{C4C4C4}
\usepackage{float}
\newcommand{\pluseq}{\mathrel{+}=}

\usepackage{stfloats}
\usepackage{mdsymbol}
\usepackage{mathrsfs} 
\usepackage{mathtools}
\usepackage{enumitem}
\usepackage{scalerel}
\newcommand\Tau{\scalerel*{\tau}{T}}

\DeclarePairedDelimiter\floor{\lfloor}{\rfloor}
\DeclarePairedDelimiter\ceil{\lceil}{\rceil}

\begin{document}
\title{On the Skew-Symmetric Binary Sequences and the Merit Factor Problem}  
\bstctlcite{IEEEexample:BSTcontrol}

\author{Miroslav Dimitrov \IEEEmembership{Student Member, IEEE}
\thanks{This work has been partially supported by the Bulgarian National Science Fund under contract number DH 12/8, 15.12.2017.}
\thanks{M. Dimitrov  is with the Institute of Mathematics and Informatics, Bulgarian Academy of Sciences, Sofia, Bulgaria (email:mirdim@math.bas.bg).}
}%

\maketitle

\begin{abstract}
The merit factor problem is of practical importance to manifold domains, such as digital communications engineering, radars, system modulation, system testing, information theory, physics, chemistry. However, the merit factor problem is referenced as one of the most difficult optimization problems and it was further conjectured that stochastic search procedures will not yield merit factors higher than 5 for long binary sequences (sequences with lengths greater than 200). Some useful mathematical properties related to the flip operation of the skew-symmetric binary sequences are presented in this work. By exploiting those properties, the memory complexity of state-of-the-art stochastic merit factor optimization algorithms could be reduced from $O(n^2)$ to $O(n)$. As a proof of concept, a lightweight stochastic algorithm was constructed, which can optimize pseudo-randomly generated skew-symmetric binary sequences with long lengths (up to ${10}^5+1$) to skew-symmetric binary sequences with a merit factor greater than 5. An approximation of the required time is also provided. The numerical experiments suggest that the algorithm is universal and could be applied to skew-symmetric binary sequences with arbitrary lengths.
\end{abstract}


\IEEEpeerreviewmaketitle

\section{Introduction}
A classical problem of digital sequence design is to determine such binary sequences whose aperiodic autocorrelation characteristics are collectively small according to some pre-defined criteria. An example of such an important measure is the merit factor. As summarized in \cite{jedwab2004survey}, Golay's publications reveal a fascination with the merit factor problem for a period of nearly twenty years. In \cite{golay1990new}, the merit factor problem was referenced by Golay as "...challenging and charming".

If $F_n$ denotes the optimal (greatest) value of the merit factor among all sequences of length $n$, then the merit factor problem could be described as finding the value of $\limsup_{n \rightarrow \infty}F_n$. Several conjectures regarding the $\limsup_{n \rightarrow \infty}F_n$ value should be mentioned. The first conjecture published in \cite{hoholdt1988determination} assumes that $\limsup_{n \rightarrow \infty}F_n = 6$. A more optimistic conjecture that $\limsup_{n \rightarrow \infty}F_n = \infty$ is given by Littlewood \cite{littlewood1966polynomials}. In \cite{byrnes1990l4}, it was conjectured that $\limsup_{n \rightarrow \infty}F_n = 5$. Golay \cite{golay1982merit} assumed that the expected value of $\limsup_{n \rightarrow \infty}F_n$ is very close to $12.32$. However, in \cite{golay1983merit} he added that "...no systematic synthesis will ever be found which will yield higher
merit factors [than 6]...". Nevertheless, in \cite{borwein2004binary} it was conjectured that $\limsup_{n \rightarrow \infty}F_n > 6.34$. The latest assumption is based on the specially constructed infinite family of sequences.

Since the merit factor problem has resisted more than 50 years of theoretical attacks, a significant number of computational pieces of evidence were collected. They could be divided into exhaustive search methods and heuristic methods. 

Regarding the exhaustive search methods, the optimal merit factors for all binary sequences with lengths $n \leq 60$ are given in \cite{mertens1996exhaustive}. Twenty years later, the list of optimal merit factors was extended to $n \leq 66$ \cite{packebusch2016low}. The two largest known values of $F_n$ are 14.1 and 12.1 for $n$ equals to respectively 13 and 11. It should be mentioned that both of those binary sequences are comprised of the Barker sequences \cite{barker1953group}. In fact, in \cite{jedwab2004survey} the author published a personal selection of challenges concerning the merit factor problem, arranged in order of increasing significance. The first suggested challenge is to find a binary sequence $X$ of length $n>13$ for which $F(X) \geq 10$.

A reasonable strategy for finding binary sequences with near-optimal merit factor is to introduce some restriction on the sequences' structure. A well-studied restriction on the structure of the sequence has been defined by the skew-symmetric binary sequences, which were introduced by Golay \cite{golay1972class}. Having a binary sequence $(b_0,b_1, \cdots, b_{2l})$ of odd length $n = 2l+1$, the restriction is defined by $b_{l+i} = (-1)^ib_{l-i} \text{ for } i=1,2,\cdots,l.$ Golay observed that odd length Barker sequences are skew-symmetric. Therefore, an idea of binary sequences' sieving was proposed \cite{golay1977sieves}. Furthermore, as shown in \cite{golay1972class}, all aperiodic autocorrelations of a skew-symmetric sequence with even indexes are equal to 0. The optimal merit factors for all skew-symmetric sequences of odd length $n \leq 59$ were given by Golay himself \cite{golay1977sieves}. Later, the optimal merit factors for skew-symmetric sequences with lengths $n \leq 69$ and $n \leq 71$ were revealed respectively in \cite{golay1990new} and \cite{de1992low}, while the optimal skew-symmetric solutions for $n \leq 89$ and $n \leq 119$ were given in respectively \cite{prestwich2013improved} and \cite{packebusch2016low}.

It should be noted, that the problem of minimizing $F_n$ is also known as "low autocorrelated binary string problem", or the LABS problem. It has been well studied in theoretical physics and chemistry. For example, the LABS problem is correlated with the quantum models of magnetism. Having this in mind, the merit factor problem was attacked by various search algorithms, such as the branch and bound algorithm proposed in \cite{packebusch2016low}, as well as stochastic search algorithms like tabu search \cite{halim2008engineering}, memetic algorithm combined with tabu search \cite{gallardo2009finding}, as well as evolutionary and genetic algorithms \cite{de1992low}\cite{militzer1998evolutionary}. However, since the search space grows like $2^n$, the difficulty of finding long binary sequences with near-optimal $F_n$ significantly increases. Bernasconi predicted that \cite{bernasconi1987low} " ... stochastic search procedures will not yield merit factors higher than about $F_n=5$  for long sequences". By long sequences, Bernasconi was referring to binary sequences with lengths greater than 200. Furthermore, in \cite{de1992low} the problem was described as " ... amongst the most difficult optimization problems".

The principle behind basic search methods could be summarized as moving through the search space by doing tiny changes inside the current binary sequence. In the case of skew-symmetric binary sequences, Golay suggested \cite{golay1975hybrid} that only one or two elements should be changed at a given optimization step. In case the new candidate has a better merit factor, the search method accepts it as a new current state and continues the optimization process. Having this in mind, a strategy of how to choose a new sequence when no acceptable neighbor sequence exists should be considered as well. 

The best results regarding skew-symmetric binary sequences with high merit factors are achieved by  \cite{gallardo2009finding}\cite{bovskovic2017low}\cite{brest2018heuristic}\cite{brest2020searching}. In \cite{gallardo2009finding}, the authors introduced a memetic algorithm with an efficient method to recompute the characteristics of a given binary sequence $L'$, such that $L'$ is one flip away from $L$, and assuming that some products of elements from $L$ have been already stored in memory. More precisely, a square $(n-1,n-1)$ tau table $\Tau(S)$, such that $\Tau(S)_{ij} = s_js_{i+j}$ for $j \leq n-i$ was introduced. Later, in \cite{bovskovic2017low} the principle of self-avoiding walk \cite{madras2013self} was considered. By using Hasse graphs the authors demonstrated that considering the LABS problem, a basic stochastic search method could be easily trapped in a cycle. To avoid this scenario, the authors suggested the usage of a self-avoiding walk strategy accompanied by a hash table for efficient memory storage of the pivot coordinates. Then, in \cite{brest2018heuristic} an algorithm called xLastovka was presented. The concept of a priority queue was introduced. In summary, during the optimization process, a queue of pivot coordinates altogether with their energy values is maintained. Recently, some skew-symmetric binary sequences with record-breaking merit factors for lengths from 301 to 401 were revealed \cite{brest2020searching}.

The aforementioned state-of-the-art algorithms are benefiting from the tau table $\Tau(S)$ previously discussed. It significantly increases the speed of evaluating a given one-flip-away neighbor, reaching a time complexity of $O(n)$. However, the memory complexity of maintaining $\Tau(S)$ is $O(n^2)$. Having this in mind, the state-of-the-art algorithms could be inapplicable to very long binary sequences due to hardware restrictions.

In this work, by using some mathematical insights, an alternative to the $\Tau(S)$ table is suggested, the usage of which significantly reduces the memory complexity of the discussed state-of-the-art algorithms from $O(n^2)$ to $O(n)$. This enhancement could be easily integrated. For example, in an online repository \cite{bovskovic2016github} a collection of currently known best merit factors for skew-symmetric sequences with lengths from 5 to 449 is given. The longest binary sequence is of length 449, having a merit factor of $6.5218$. As a proof of concept, by using just a single budget processor Xeon-2640 CPU with a base frequency of 2.50 GHz, the price of which at the time of writing this paper is about 15 dollars, and our tweaked implementation of the lssOrel algorithm introduced in \cite{bovskovic2016github}, we were able to find a skew-symmetric binary sequence with better merit factor of $6.5319$. The time required was approximately one day. As a comparison, the currently known optimal results were acquired by using the Slovenian Initiative for National Grid (SLING) infrastructure (100 processors) and a 4-day threshold limitation per length.

It should be noted, that despite the significant memory complexity optimization introduced with the current paper, the state-of-the-art algorithms could still suffer from memory and speed issues. As previously discussed, additional memory-requiring structures were needed, such as, for example, a set of all previously visited pivots \cite{bovskovic2017low} or a priority queue with 640 000 coordinates and a total size of 512MB \cite{brest2018heuristic}. 

Another issue is the "greedy" approach of collecting all the neighbors to determine the best one. This could dramatically decrease the optimization process, especially when very long binary sequences are involved. This side-effect is further discussed in \cite{dimitrov2020generation}.

Having those observations in mind, an almost memory-free optimization algorithm is suggested. More precisely, both the time and memory complexities of the algorithm are linear. This could be particularly beneficial for multi-thread architectures or graphical processing units. During our experiments, and by using the aforementioned algorithm, we were able to find skew-symmetric sequences with merit factors strictly greater than $F_n = 5$ for all the tested lengths up to $10^5+1$. Thus, Bernasconi's prediction that no stochastic search procedure will yield merit factors higher than $F_n = 5$ for binary sequences with lengths greater than 200 was very pessimistic. 

\section{Preliminaries}
\label{sec:prelims}

We denote as $B=(b_0,b_1,\cdots ,b_{n-1})$ the binary sequence with length $n>1$, such that $b_i\in \{-1,1\}, 0\leq i\leq n-1$. The aperiodic autocorrelation function of $B$ is given by $C_u(B)=\sum_{j=0}^{n-u-1} b_jb_{j+u}, \ \ for \ u\in \{0,1,\cdots, n-1\}.$ We define $C_u(B)$ for $u\in \{1, \cdots ,n-1\}$ as a sidelobe level. $C_0(B)$ is called the mainlobe. We define the peak sidelobe level, or PSL, of $B$ as $B_{PSL}=\max_{0<u<n} \lvert C_u(B)\rvert.$ The enery $\mathbb{E}(B)$ of $B$ is defined as $\mathbb{E}(B) = \sum_{u=1}^{n-1}C_u(B)^2,$ while the merit factor, or \textbf{MF}, $\mathbb{MF}(B)$ of $B$ is defined as $\mathbb{MF}(B) = \frac{n^2}{2\mathbb{E}(B)}.$ Let us denote $C_{n-i-1}(B)$ by $\hat{C}_i(B)$. Since this is just a rearrangement of the sidelobes of $B$, it follows that $B_{PSL}=\max_{0<u<n} \lvert C_u(B)\rvert = \max_{0 \le u < n-1} \lvert \hat{C}_u(B)\rvert.$

\section{Skew-Symmetric Sequences and their MF}
\label{sec:problemRevisited}

Let us consider a skew-symmetric binary sequence defined by an array $L = \left[b_0, b_1, \cdots, b_{n-1}\right]$ with an odd length $n =2l+1$.  If the corresponding to $L$ sidelobes' array is denoted by an array $W$, we have: $W= \left[C_{n-1}(L), C_{n-2}(L), \cdots, C_{1}(L), C_{0}(L)\right],$ where $C_u(L)=\sum_{j=0}^{n-u-1} b_jb_{j+u}, \ \ for \ u\in \{0,1,\cdots, n-1\}.$

In this paper, for convenience, we will use the reversed version of $W$, denoted by $S$, s.t:
$$S= \left[\hat{C}_{0}(L), \hat{C}_{1}(L), \cdots, \hat{C}_{n-2}(L), \hat{C}_{n-1}(L)\right],$$
where $\hat{C}_{n-i-1}(L) = C_i(L)$, for $i \in \lbrace 0,1,\cdots,n-1\rbrace$. Thus, $\hat{C}_i(L) = C_{n-i-1}(L) = \sum_{j=0}^{n-(n-i-1)-1} b_jb_{j+(n-i-1)}.$

Hence, $\hat{C}_i(L)= \sum_{j=0}^{i} b_jb_{j+n-i-1}, \ \ for \ i\in \{0,1,\cdots, n-1\} .$

Furthermore, we will denote the $i$-th element of a given array $A$ as $A[i]$. It should be noted that the first index of an array is 0, not 1. For example, $W[n-1] = S[0] = \hat{C}_0[L] = C_{n-1}(L).$

Since $L$ is a skew-symmetric binary sequence, the following properties hold:

\begin{itemize}
\item{$S[i]=0$, for odd values of $i$.}
\item{$L[l-i] = {(-1)}^iL[l+i]$}
\end{itemize}
Having this in mind, the array of sidelobes $S$ could be represented as follows: $$S= \left[\hat{C}_{0}(L), 0, \hat{C}_{2}(L),0, \cdots, 0, \hat{C}_{n-3}(L), 0, \hat{C}_{n-1}(L)\right].$$

For convenience, we will use the notation $S_i$ which represents the $(i-1)$-th element of a given array S, or more formally $S_i = S[i-1]$. Thus, for every odd value $r$, we have $S_r = \hat{C}_{r-1}(L) = \sum_{j=0}^{r-1} b_jb_{j+n-r+1-1} = \sum_{j=0}^{r-1} b_jb_{j+n-r} = \sum_{j=1}^{r} b_{j-1}b_{j-1+n-r}.$

In terms of $L$, the previous relationship could be written down as follows: $$S_r = \sum_{j=1}^{r} b_{j-1}b_{j-1+n-r} =  \sum_{i=1}^{r}L[i-1]L[n+i-r-1]$$

We could further substitute $i=l-q$, for $q \in \{0,1,\cdots,l\}$ into the major property of the skew-symmetric sequences to show that: $L[l-l+q] = {(-1)}^{l-q}L[l+l-q] \implies $
$L[q] = {(-1)}^{l-q}L[l+l+1-q-1] \implies $
$L[q] = {(-1)}^{l-q}L[n-q-1]. $

Hence, given a skew-symmetric sequence $L$ with length $n=2l+1$, if we flip both the elements on positions $q$ and $n-q-1$, for some fixed $q \in \{0,1,\cdots,l\}$, the resulted binary sequence $L^q$ will be skew-symmetric as well. Let's denote the array of sidelobes of $L^q$ as $S^q$, i.e:
$$S^q= \left[\hat{C}_{0}(L^q), 0, \hat{C}_{2}(L^q),0, \cdots, 0, \hat{C}_{n-3}(L^q), 0, \hat{C}_{n-1}(L^q)\right].$$

By a consequence of the previously aforementioned observations, we have $S^q_r =  \sum_{i=1}^{r}L^q[i-1]L^q[n+i-r-1].$

In Theorem \ref{theorem:transformation} a more detailed picture of the $S$ array transformation to the $S^q$ array is provided.

\begin{theorem}
\label{theorem:transformation}
Given two skew-symmetric sequences $L$ and $L^q$ with length $n=2l+1$, and with sidelobes arrays respectively $S$ and $S^q$, where $q<l$, the following properties hold:
\begin{enumerate}[label=\Roman*]
\item{For $\forall e$, s.t. $e$ is an even number, $S^q_e - S_e = 0$  }
\item{If $r$ is an odd number and $r \leq q$, $S^q_r - S_r = 0$. }
\item{If $r$ is an odd number and $r>q$, and $r<n-q$, and $q \neq r-q-1$, then: $$S^q_r - S_r = -2\left( L[q]L[n+q-r]+L[r-q-1]L[n-q-1]\right).$$}
\item{If $r$ is an odd number and $r>q$, and $r<n-q$, and $q = r-q-1$, then $S^q_r - S_r = 0$}
\item{If $r$ is an odd number and $r \geq n-q$, and $q \neq r-q-1$, then $$S^q_r - S_r = -2L[n-q-1]L[2n-q-r-1] -2L[q+r-n]L[q] - 2L[q]L[n+q-r]-2L[r-q-1]L[n-q-1]$$}
\item{If $r$ is an odd number and $r \geq n-q$, and $q = r-q-1$, then $$S^q_r - S_r = -2L[n-q-1]L[2n-q-r-1] -2L[q+r-n]L[q] $$}
\end{enumerate}
\end{theorem}

\begin{IEEEproof}
\textbf{Property I}: For $\forall e$, s.t. $e$ is an even number, $S_e = 0$ and $S^q_e= 0$, since both $S$ and $S^q$ are skew-symmetric sequences. Therefore, $S^q_e - S_e = 0$.

\textbf{Property II}: If $r$ is an odd number and $r \leq q$, then $$S^q_r - S_r = \sum_{i=1}^{r}L^q[i-1]L^q[n+i-r-1] - \sum_{i=1}^{r}L[i-1]L[n+i-r-1].$$

By construction, $L^q[q] \neq L[q]$, $L^q[n-q-1] \neq L[n-q-1]$ and $\forall x \in [0,1,\cdots ,n-1 ], x \neq q \ \&\  x \neq n-q-1 : L^q[x] = L[x]$. Since, by considering the initial condition $r \leq q$, it follows that $r-1 < q$. Therefore, for $i \in \{1,2,\cdots,r\}$,  $i-1 \leq r-1 < q$ and $L^q[i-1]=L[i-1]$. On the other hand, for $i \in \{1,2,\cdots,r\}$, $n+i-r-1 \geq n+1-r-1 = n-r$, but since $r \leq q$, then $n-r \geq n-q > n-q-1$, which means that $L^q[n+i-r-1]=L[n+i-r-1]$.

By combining the aforementioned observations:
\begin{equation} \label{theorem1:property2}
\begin{split}
 & S^q_r - S_r = \sum_{i=1}^{r}L^q[i-1]L^q[n+i-r-1] - \sum_{i=1}^{r}L[i-1]L[n+i-r-1] =\\
 & = \sum_{i=1}^{r}L[i-1]L[n+i-r-1] - \sum_{i=1}^{r}L[i-1]L[n+i-r-1] = 0 
\end{split}
\end{equation}

\textbf{Property III} We consider $r$ as an odd number, $r>q$, $r<n-q$, and $q \neq r-q-1$. Since $r>q$,  we have $r-1 \geq q$, which means that at least one element from the elements defined by $L^q[i-1]$, for $i \in \{1,2,\cdots,r\}$, will coincide with $L^q[q]$. However, since $r < n-q$, or $r-1 < n-q-1$, there will be no element from the elements defined by $L^q[i-1]$, for $i \in \{1,2,\cdots,r\}$, that will coincide with $L^q[n-q-1]$.

For $i \in \{1,2,\cdots,r\}$, $n+i-r-1 \geq n-r$.  If $n-r \leq q$ then $n-q \leq r$, which contradicts the initial condition of $r < n-q$. Therefore, $n-r > q$ and $n+i-r-1 > q$, and there will be no element from the elements defined by $L^q[n+i-r-1]$, for $i \in \{1,2,\cdots,r\}$, that will coincide with $L^q[q]$. On the other hand, for $i \in \{1,2,\cdots,r\}$, $n+i-r-1 \geq n-r$, and since $r>q$, $n-r < n-q$. Thus $n-r \leq n-q-1$, which means there will be an element from the elements defined by $L^q[n+i-r-1]$, for $i \in \{1,2,\cdots,r\}$, which will coincide with $L^q[n-q-1]$.
By combining the aforementioned observations, we have:
\begin{equation} 
\label{theorem1:property3}
\begin{split}
 & S^q_r - S_r = \sum_{i=1}^{r}L^q[i-1]L^q[n+i-r-1] - \sum_{i=1}^{r}L[i-1]L[n+i-r-1]=\\
 & = (\sum_{i=1}^{q}L^q[i-1]L^q[n+i-r-1]) + L^q[q]L^q[n+q-r] + (\sum_{i=q+2}^{r}L^q[i-1]L^q[n+i-r-1]) - \\ 
 & - (\sum_{i=1}^{q}L[i-1]L[n+i-r-1]) - L[q]L[n+q-r] - \sum_{i=q+2}^{r}L[i-1]L[n+i-r-1] \\ 
\end{split}
\end{equation}

However, since it is given that $q \neq r-q-1$, then $n+q-r \neq n+r-q-1-r = n-q-1$. Thus, the coincide elements are still to be determined inside the sequences defined for $i \in \{q+2, q+3, \cdots, r\}$. Furthermore, as previously shown, we have:

$$\sum_{i=1}^{q}L^q[i-1]L^q[n+i-r-1] = \sum_{i=1}^{q}L[i-1]L[n+i-r-1],$$

Hence:
\begin{equation} \label{theorem1:property3}
\begin{split}
& S^q_r - S_r = L^q[q]L^q[n+q-r] + (\sum_{i=q+2}^{r}L^q[i-1]L^q[n+i-r-1]) - L[q]L[n+q-r] - \sum_{i=q+2}^{r}L[i-1]L[n+i-r-1] = \\
& = L^q[q]L^q[n+q-r] + (\sum_{i=q+2}^{r-q-1}L^q[i-1]L^q[n+i-r-1]) + L^q[r-q-1]L^q[n+r-q-r-1] + \\
& + (\sum_{i=r-q+1}^{r}L^q[i-1]L^q[n+i-r-1]) - L[q]L[n+q-r] - \sum_{i=q+2}^{r-q-1}L[i-1]L[n+i-r-1] -\\
& - L[r-q-1]L[n+r-q-r-1] - \sum_{i=r-q+1}^{r}L[i-1]L[n+i-r-1]\\ 
\end{split} 
\end{equation}

Since we have isolated all coincidences, it follows:

$$\sum_{i=q+2}^{r-q-1}L^q[i-1]L^q[n+i-r-1] = \sum_{i=q+2}^{r-q-1}L[i-1]L[n+i-r-1]$$
$$\sum_{i=r-q+1}^{r}L^q[i-1]L^q[n+i-r-1] = \sum_{i=r-q+1}^{r}L[i-1]L[n+i-r-1]$$

Thus, 
\begin{equation} \label{theorem1:property3}
\begin{split}
& S^q_r - S_r = L^q[q]L^q[n+q-r] + L^q[r-q-1]L^q[n-q-1] - L[q]L[n+q-r] - L[r-q-1]L[n-q-1]\\
\end{split} 
\end{equation}

However, since $L^q$ is identical to $L$ with $q$-th and $n-q-1$-th bits flipped, we have $L^q[q] = -L[q]$ and $L^q[n-q-1] = -L[n-q-1]$. 

\begin{equation} \label{theorem1:property3}
\begin{split}
& S^q_r - S_r = - L[q]L^q[n+q-r] - L^q[r-q-1]L[n-q-1] - L[q]L[n+q-r] - L[r-q-1]L[n-q-1] = \\
& = - L[q]L[n+q-r] - L[r-q-1]L[n-q-1] -   L[q]L[n+q-r] - L[r-q-1]L[n-q-1] = \\
& = -2\left(L[q]L[n+q-r] + L[r-q-1]L[n-q-1] \right) \\
\end{split} 
\end{equation}

\textbf{Property IV} This property is almost identical to Property III. However, this time the fact that $q = r-q-1$ should be considered. More precisely, we should revisit the equation:

\begin{equation}
\begin{split}
& S^q_r - S_r = L^q[q]L^q[n+q-r] + (\sum_{i=q+2}^{r}L^q[i-1]L^q[n+i-r-1]) - L[q]L[n+q-r] - \sum_{i=q+2}^{r}L[i-1]L[n+i-r-1] \\
\end{split}
\end{equation}

Since $q = r-q-1$, or $2q=r-1$, and $n+q-r = n+q-2q-1 = n-q-1$, both coincides appeared on the same monomial:
$$\sum_{i=q+2}^{r}L^q[i-1]L^q[n+i-r-1] = \sum_{i=q+2}^{r}L[i-1]L[n+i-r-1]$$

Therefore, 
\begin{equation}
\begin{split}
& S^q_r - S_r = L^q[q]L^q[n+q-r] - L[q]L[n+q-r] = L^q[q]L^q[n-q-1] - L[q]L[n-q-1] = \\
& = - L[q]L^q[n-q-1] - L[q]L[n-q-1] = L[q]L[n-q-1] - L[q]L[n-q-1] = 0 \\
\end{split}
\end{equation}

\textbf{Property V} We have that $r \geq n-q$, while in the same time $q \neq r-q-1$. We continue the proof of this and consequence properties by following the same method and observations made throughout the proof of Properties III and IV. A total of 4 coincides between $L^q$ and $L$ are possible:

\begin{itemize}
\item{$i-1=q$, or $i=q+1$}
\item{$n+i-r-1=q$, or $i=q+r-n+1$}
\item{$i-1=n-q-1$, or $i=n-q$}
\item{$n+i-r-1=n-q-1$, or $i=r-q$}
\end{itemize}

\begin{equation} 
\label{theorem1:property3}
\begin{split}
 & S^q_r - S_r = \sum_{i=1}^{r}L^q[i-1]L^q[n+i-r-1] - \sum_{i=1}^{r}L[i-1]L[n+i-r-1] = \\
 & = \sum_{i=1, i \not\in \{ q+1, q+r-n+1,n-q,r-q\}}^{r}L^q[i-1]L^q[n+i-r-1] + L^q[q]L^q[n+q-r] + L^q[q+r-n]L^q[q] + \\
 & + L^q[n-q-1]L^q[2n-q-r-1] + L^q[r-q-1]L^q[n-q-1] - \sum_{i=1, i \not\in \{ q+1, q+r-n+1,n-q,r-q\}}^{r}L[i-1]L[n+i-r-1] - \\ 
 & - L[q]L[n+q-r] - L[q+r-n]L[q] - L[n-q-1]L[2n-q-r-1] -  L[r-q-1]L[n-q-1]\\
\end{split}
\end{equation}

Since $L^q$ is identical to $L$ with $q$-th and $n-q-1$-th bits flipped, it follows that both sums are comprised of non-flipped bits, and therefore they are equal. Thus:
\begin{equation} 
\label{theorem1:property3}
\begin{split}
 & S^q_r - S_r = L^q[q]L^q[n+q-r] + L^q[q+r-n]L^q[q] + L^q[n-q-1]L^q[2n-q-r-1] + L^q[r-q-1]L^q[n-q-1] - \\
 & - L[q]L[n+q-r] - L[q+r-n]L[q] - L[n-q-1]L[2n-q-r-1] - L[r-q-1]L[n-q-1] = \\
 & = -L[q]L[n+q-r] - L[q+r-n]L[q] - L[n-q-1]L[2n-q-r-1] - L[r-q-1]L[n-q-1] - \\
 & - L[q]L[n+q-r] - L[q+r-n]L[q] - L[n-q-1]L[2n-q-r-1] - L[r-q-1]L[n-q-1] = \\ 
 & = -2*(L[q]L[n+q-r] + L[q+r-n]L[q] + L[n-q-1]L[2n-q-r-1] + L[r-q-1]L[n-q-1])\\
\end{split}
\end{equation}

\textbf{Property VI} This property is very similar to the previous Property V. However, since $q = r-q-1$, and by using the similar approach shown throughout the proof of Property IV, we could exactly pinpoint   those monomials that include a double coincide. Indeed, when $q = r-q-1$, $n+q-r = n + (r-q-1) - r = n-q-1$. Thus:
\begin{equation} 
\label{theorem1:property3}
\begin{split}
 & S^q_r - S_r = \sum_{i=1}^{r}L^q[i-1]L^q[n+i-r-1] - \sum_{i=1}^{r}L[i-1]L[n+i-r-1] = \\
 & = \sum_{i=1, i \not\in \{ q+1, q+r-n+1,n-q,r-q\}}^{r}L^q[i-1]L^q[n+i-r-1] + L^q[q]L^q[n+q-r] + L^q[q+r-n]L^q[q] + \\
 & + L^q[n-q-1]L^q[2n-q-r-1] - \sum_{i=1, i \not\in \{ q+1, q+r-n+1,n-q,r-q\}}^{r}L[i-1]L[n+i-r-1] - \\
 & - L[q]L[n+q-r] - L[q+r-n]L[q] - L[n-q-1]L[2n-q-r-1] \\
\end{split}
\end{equation}

However:
\begin{equation} 
\label{theorem1:property3}
\begin{split}
& L^q[q]L^q[n+q-r] - L[q]L[n+q-r] = L^q[q]L^q[n-q-1] - L[q]L[n-q-1] =\\
& = (-1)L[q](-1)L[n-q-1] - L[q]L[n-q-1] = 0 \\
\end{split}
\end{equation}
Thus:
\begin{equation} 
\label{theorem1:property3}
\begin{split}
 & S^q_r - S_r = \sum_{i=1, i \not\in \{ q+1, q+r-n+1,n-q,r-q\}}^{r}L^q[i-1]L^q[n+i-r-1] + L^q[q+r-n]L^q[q] + \\
 & + L^q[n-q-1]L^q[2n-q-r-1] - \sum_{i=1, i \not\in \{ q+1, q+r-n+1,n-q,r-q\}}^{r}L[i-1]L[n+i-r-1] - \\
 & - L[q+r-n]L[q] - L[n-q-1]L[2n-q-r-1]\\
\end{split}
\end{equation}

Following the same observations made throughout the proof of Property V, the equation could be further simplified to:
\begin{equation} 
\label{theorem1:property3}
\begin{split}
 & S^q_r - S_r = -2*(L[q+r-n]L[q] + L[n-q-1]L[2n-q-r-1])\\
\end{split}
\end{equation}
\end{IEEEproof}

We should emphasize, that Theorem \ref{theorem:transformation} covers all the possible sidelobes positions and all the possible flip bit choices. Indeed, let's define the sidelobe position as $s$, while the flip bit position as $q$. Furthermore, we denote property $X$ as $\delta_X$. Then:
\begin{equation} 
\label{theorem1:property3}
\begin{split}
& \forall s \forall q \equiv (\forall e: e\equiv0 \bmod 2) \forall q  \bigcup (\forall r: r\equiv1 \bmod 2) \forall q \equiv \delta_1 \bigcup (\forall r: r\equiv1 \bmod 2)(\forall q : r \leq q) \bigcup \\
& \bigcup (\forall r: r\equiv1 \bmod 2)(\forall q : r > q) = \delta_1 \bigcup \delta_2 \bigcup (\forall r: r\equiv1 \bmod 2)(\forall q : r > q, r < n-q) \bigcup \\
& \bigcup (\forall r: r\equiv1 \bmod 2)(\forall q : r > q, r \geq n-q)
\end{split}
\end{equation}

For convenience, we will substitute $(\forall r: r\equiv1 \bmod 2)$ as $\forall r \in \mathbb{O}$:
\begin{equation} 
\label{theorem1:property3}
\begin{split}
& \delta_1 \bigcup \delta_2 \bigcup (\forall r \in \mathbb{O})(\forall q : r > q, r < n-q) \bigcup (\forall r \in \mathbb{O})(\forall q : r > q, r \geq n-q) = \\ 
& = \delta_1 \bigcup \delta_2 \bigcup (\forall r \in \mathbb{O})(\forall q : r > q, r < n-q, q \neq r-q-1) \bigcup (\forall r \in \mathbb{O})(\forall q : r > q, r < n-q, q = r-q-1) \bigcup \\
& \bigcup (\forall r \in \mathbb{O})(r \geq n-q) = \bigcup_{i=1}^4\delta_i \bigcup (\forall r \in \mathbb{O})(r \geq n-q) = \bigcup_{i=1}^4\delta_i \bigcup (\forall r \in \mathbb{O})(r \geq n-q, q \neq r-q-1) \bigcup \\
& \bigcup (\forall r \in \mathbb{O})(r \geq n-q, q = r-q-1) = \bigcup_{i=1}^6\delta_i
\end{split} 
\end{equation}

Furthermore, $\bigcap_{i=1}^6\delta_i = \{ \O \}$. Theorem \ref{theorem:transformation}, as well as the observations made throughout this section, are summarized as a pseudo-code in Algorithm \ref{algor:InMemoryFlip}. The following notations were used:

\begin{itemize}
\item{$n=2l+1$: the odd length of the sequence }
\item{$q$: the bit position which is to be flipped. Defined for $q<l$. Please note, that besides $q$, the algorithm is going to flip $n-q-1$ as well, since we want to keep the skew-symmetric property of the binary sequence. }
\item{$L$: a binary skew-symmetric sequence}
\item{$S$: the sidelobes array corresponding to $L$}
\end{itemize}

\algrenewcommand\algorithmicindent{0.5em}%
\begin{algorithm}[]
\caption{The in-memory flip of skew-symmetric binary sequence in linear time and memory complexities}
\label{algor:InMemoryFlip}
\begin{algorithmic}[1]
\Procedure{Flip}{$q, L, S$}
\For{$r =1; r < n-1; r += 2$}
		\If{$r \leq q$}
			\State \textbf{continue}
		\EndIf
		\State $\epsilon_1 = L[q], \epsilon_2 = L[n+q-r], \epsilon_3 = L[r-q-1] $
		\State $\epsilon_4 = L[n-q-1], \epsilon_5 = L[2n-q-r-1], \epsilon_6 = L[q+r-n]$
		\If{$r < n-q$}
			\If{$q \neq r-q-1$}
				\State $S_r = S_r -  2(\epsilon_1\epsilon_2 + \epsilon_3\epsilon_4)$	
			\EndIf		
		\Else
			\If{$q \neq r-q-1$}
				\State $S_r = S_r -  2(\epsilon_1\epsilon_2 + \epsilon_3\epsilon_4 + \epsilon_4\epsilon_5 + \epsilon_6\epsilon_1)$		
			\Else 
				\State $S_r = S_r -  2(\epsilon_4\epsilon_5 + \epsilon_6\epsilon_1)$
			\EndIf
		\EndIf
	\EndFor
	\State $L[q] = -L[q], L[n-q-1] = -L[n-q-1]$
\EndProcedure
\end{algorithmic}
\end{algorithm} 

When the algorithm finishes, $L$ is going to be modified to $L^q$, while $S$ is going to correspond to the sidelobes array of $L^q$. This is accomplished in $O(n)$ for both time and memory complexities.

\section{The Energy and Merit Factor of $L^q$}
\label{sec:fitness}

\begin{theorem}
\label{theorem:meritFactor}
Given two skew-symmetric sequences $L$ and $L^q$ with length $n=2l+1$, where $L^q$ corresponds to $L$ with $q$-th and $n-q-1$-th bit flipped for some fixed $q<l$, and with sidelobes arrays denoted respectively as $S$ and $S^q$, the following property holds:
\begin{equation} 
\label{theorem1:property3}
\begin{split}
& \mathbb{E}(L^q) = \mathbb{E}(L) + \sum_{r=q+1, r \neq 2q+1}^{n-q-1}(16+\sigma\kappa\epsilon_1) + \sum_{r=n-q, r \neq 2q+1}^{n-1} (\kappa(\epsilon_2+\sigma\epsilon_1) + 32 + 32\sigma\epsilon_1\epsilon_2) + \sum_{r \geq n-q, r \leq n-1, r = 2q+1}(16+\kappa\epsilon_2), \\
\end{split}
\end{equation}
where $\sigma=(-1)^{l-q}$, $\kappa = -8S_rL[q]$, $\epsilon_1(r) = L[r-q-1]$, $\epsilon_2(r) = L[q+r-n]$.
\end{theorem}

\begin{IEEEproof}

\begin{equation} 
\label{theorem1:property3}
\begin{split}
& \mathbb{E}(L^q) - \mathbb{E}(L) = \sum_{i=1}^{n-1}{(S_i^q)}^2 - \sum_{i=1}^{n-1}{(S_i)}^2 = \sum_{i=1}^{n-1}{((S_i^q)^2 - (S_i)}^2) = \\
& = \sum_{j=1}^{6}\sum_{i \in \mathbb{D}(\delta_j)}{((S_i^q)^2 - (S_i)}^2) = \sum_{j=1}^{6}\sum_{i \in \mathbb{D}(\delta_j)}{((S_i + \delta_j)^2 - (S_i)}^2) = \sum_{j=1}^{6}\sum_{i \in \mathbb{D}(\delta_j)}{(2S_i\delta_j + \delta_j^2) }\\
\end{split}
\end{equation}

We proceed with the calculation of $\delta_i^2$, for $i \in [3,5,6]$. 
\begin{equation} 
\label{theorem1:property3}
\begin{split}
& \delta_3^2 = {\left( -2\left( L[q]L[n+q-r]+L[r-q-1]L[n-q-1]\right) \right)}^2 = \\
& = 4(L[q]^2L[n+q-r]^2 + L[r-q-1]^2L[n-q-1]^2 + 2L[q]L[n+q-r]L[r-q-1]L[n-q-1]) \\ 
\end{split}
\end{equation}
However, $L[x]^2 = 1$ for any x, therefore:
\begin{equation} 
\label{theorem1:property3}
\begin{split}
& \delta_3^2 = 4(1 + 1 + 2L[q]L[n+q-r]L[r-q-1]L[n-q-1])  \\
\end{split}
\end{equation}

Furthermore, from the main property of the skew-symmetric binary sequences, we know that $L[q] = {(-1)}^{l-q}L[n-q-1]$:
\begin{equation} 
\label{theorem1:property3}
\begin{split}
& L[n+q-r] = {(-1)}^{l-(n+q-r)}L[n-(n+q-r)-1] =  {(-1)}^{l-n-q+r)}L[n-n-q+r-1] = {(-1)}^{l-n-q+r)}L[r-q-1] \\
\end{split}
\end{equation}

However, since $r \equiv n \equiv 1 \bmod 2$, we know that $r-n \equiv 0 \bmod 2$ and therefore ${(-1)}^{l-n-q+r} = {(-1)}^{l-q}$. Having this in mind, we can further simplify $\delta_3^2$:

\begin{equation} 
\label{theorem1:property3}
\begin{split}
& \delta_3^2 = 8 + 8L[q]L[n+q-r]L[r-q-1]L[n-q-1]) = 8 + 8L[q]{(-1)}^{l-q}L[r-q-1])L[r-q-1]{(-1)}^{l-q}L[q] = \\
& = 8 + 8L[q]^2{(-1)}^{2(l-q)}L[r-q-1]^2 =  8+8 = 16\\
\end{split}
\end{equation}

\begin{equation} 
\label{theorem1:property3}
\begin{split}
& \delta_5^2 = (-2L[n-q-1]L[2n-q-r-1] -2L[q+r-n]L[q] - 2L[q]L[n+q-r]-2L[r-q-1]L[n-q-1]))^2 \\
\end{split}
\end{equation}
We could simplify $L[2n-q-r-1]$:
\begin{equation} 
\label{theorem1:property3}
\begin{split}
& L[2n-q-r-1] = {(-1)}^{l-(2n-q-r-1)}L[n-(2n-q-r-1)-1] =  {(-1)}^{l-2n+q+r+1)}L[-n+q+r] \\
\end{split}
\end{equation}

Since $r$ is odd, $r+1$ is even, and therefore $r+1-2n \equiv 0 \bmod 2$. Therefore, ${(-1)}^{l-2n+q+r+1)} = {(-1)}^{l+q} = {(-1)}^{l-q}{(-1)}^{2q} = {(-1)}^{l-q}$:
\begin{equation} 
\label{theorem1:property3}
\begin{split}
& \delta_5^2 = 4(L[n-q-1]L[2n-q-r-1] + L[q+r-n]L[q] + L[q]L[n+q-r] + L[r-q-1]L[n-q-1]))^2 = \\
& = 4({(-1)}^{l-q}L[q]{(-1)}^{l-q}L[r+q-n] + L[q+r-n]L[q] + L[q]{(-1)}^{l-q}L[r-q-1] + L[r-q-1]{(-1)}^{l-q}L[q])^2 = \\
& = 4(2L[q]L[r+q-n] + 2L[q]L[r-q-1]{(-1)}^{l-q})^2 = 16L[q]^2(L[r+q-n]+L[r-q-1]{(-1)}^{l-q})^2 = \\
& = 16(L[r+q-n]^2 + (L[r-q-1]{(-1)}^{l-q})^2 + 2L[r+q-n]L[r-q-1]{(-1)}^{l-q}) = \\
& = 32 + 32L[r+q-n]L[r-q-1]{(-1)}^{l-q}
\end{split}
\end{equation}
Finally, we simplify $\delta_6^2$:
\begin{equation} 
\label{theorem1:property3}
\begin{split}
& \delta_6^2 = 4(L[n-q-1]L[2n-q-r-1] + L[q+r-n]L[q])^2 = 4(L[n-q-1]^2L[2n-q-r-1]^2 + L[q+r-n]^2L[q]^2 + \\
& + 2L[n-q-1]L[2n-q-r-1]L[q+r-n]L[q]) = 4(2+2{(-1)}^{l-q}L[q]{(-1)}^{l-q}L[r+q-n]L[q+r-n]L[q] = \\
& = 4(2+2{(-1)}^{2(l-q)}L[q]^2L[q+r-n]^2) = 16
\end{split}
\end{equation}
We have:
\begin{equation} 
\label{theorem1:property3}
\begin{split}
& \mathbb{E}(L^q) - \mathbb{E}(L) = \sum_{j=1}^{6}\sum_{i \in \mathbb{D}(\delta_j)}{(2S_i\delta_j + \delta_j^2)} = \sum_{j \in \{1,2,4\}}\sum_{i \in \mathbb{D}(\delta_j)}{(2S_i\delta_j + \delta_j^2)} + \sum_{j \in \{3,5,6\}}\sum_{i \in \mathbb{D}(\delta_j)}{(2S_i\delta_j + \delta_j^2)} \\
& = 
\end{split}
\end{equation}
However, since $\delta_j$, for $j \in \{1,2,4\}$ is 0:
\begin{equation} 
\label{theorem1:property3}
\begin{split}
& \mathbb{E}(L^q) - \mathbb{E}(L) =  \sum_{j \in \{3,5,6\}}\sum_{i \in \mathbb{D}(\delta_j)}{(2S_i\delta_j + \delta_j^2)} = \\
& = \sum_{r=q+1, r \neq 2q+1}^{n-q-1}(2S_r\delta_3 + \delta_3^2) + \sum_{r=n-q, r \neq 2q+1}^{n-1}(2S_r\delta_5 + \delta_5^2) + \sum_{r \geq n-q, r \leq n-1, r = 2q+1}(2S_r\delta_6 + \delta_6^2) \\
& + 
\end{split}
\end{equation}

Moreover, since:
\begin{equation} 
\label{theorem1:property3}
\begin{split}
& \delta_3 = -2\left( L[q]L[n+q-r]+L[r-q-1]L[n-q-1]\right) = -2\left( L[q]{(-1)}^{l-q}L[r-q-1]+L[r-q-1]{(-1)}^{l-q}L[q]\right) = \\
& = -4{(-1)}^{l-q}L[q]L[r-q-1] = -4\sigma L[q]\epsilon_1
\end{split}
\end{equation}

\begin{equation} 
\label{theorem1:property3}
\begin{split}
& \delta_5 = -2(L[n-q-1]L[2n-q-r-1] + L[q+r-n]L[q] +  L[q]L[n+q-r] + L[r-q-1]L[n-q-1]) = \\
& = -4(L[q]L[r+q-n] + L[q]L[r-q-1]{(-1)}^{l-q}) = -4L[q](L[r+q-n] + L[r-q-1]{(-1)}^{l-q}) = -4L[q](\epsilon_2 + \epsilon_1\sigma) \\
\end{split}
\end{equation}

\begin{equation} 
\label{theorem1:property3}
\begin{split}
& \delta_6 = -2(L[n-q-1]L[2n-q-r-1] + L[q+r-n]L[q]) = -2({(-1)}^{l-q}L[q]{(-1)}^{l-q}L[q+r-n] + L[q+r-n]L[q]) = \\
& = -4{(-1)}^{l-q}L[q]L[q+r-n] = -4\sigma L[q]\epsilon_2
\end{split}
\end{equation}

we could substitute and further simplify the difference between the merit factors of $L^q$ and $L$, i.e:
\begin{equation} 
\label{theorem1:property3}
\begin{split}
& \mathbb{E}(L^q) - \mathbb{E}(L) = \sum_{r=q+1, r \neq 2q+1}^{n-q-1}(2S_r(-4\sigma L[q]\epsilon_1) + 16) + \sum_{r=n-q, r \neq 2q+1}^{n-1}(2S_r(-4L[q](\epsilon_2 + \epsilon_1\sigma)) + 32 + 32\epsilon_2\epsilon_1\sigma) + \\
& + \sum_{r \geq n-q, r \leq n-1, r = 2q+1}(2S_r(-4\sigma L[q]\epsilon_2) + 16)
\end{split}
\end{equation}
We denote $\kappa$ as $\kappa = -8S_rL[q]$:
\begin{equation} 
\label{theorem1:property3}
\begin{split}
& \mathbb{E}(L^q) - \mathbb{E}(L) = \sum_{r=q+1, r \neq 2q+1}^{n-q-1}(-8S_r\sigma L[q]\epsilon_1 + 16) + \sum_{r=n-q, r \neq 2q+1}^{n-1}(-8S_rL[q](\epsilon_2 + \epsilon_1\sigma)) + 32 + 32\epsilon_2\epsilon_1\sigma) + \\
& + \sum_{r \geq n-q, r \leq n-1, r = 2q+1}(-8S_r\sigma L[q]\epsilon_2 + 16) = \sum_{r=q+1, r \neq 2q+1}^{n-q-1}(\kappa\sigma\epsilon_1 + 16) + \sum_{r=n-q, r \neq 2q+1}^{n-1}(\kappa(\epsilon_2 + \epsilon_1\sigma)) + 32 + 32\epsilon_2\epsilon_1\sigma) +\\
& + \sum_{r \geq n-q, r \leq n-1, r = 2q+1}(\kappa\sigma\epsilon_2 + 16)
\end{split}
\end{equation}
\end{IEEEproof}

In Algorithm \ref{algor:InMemoryProbing} a pseudo-code of the derivative function is given. The input of the function consists of a bit position $q$ to be flipped, a skew-symmetric sequence $L$ with an odd length $n=2l+1$, as well as the corresponding sidelobe array $S$. We recall that besides the bit position $q$, s.t. $q<l$, the bit position $n-q-1$ is flipped as well, so to keep the skew-symmetric property of the binary sequence. The output of the function consists of a single integer value $\Delta$, which corresponds to the difference of the energies of $L$ and $L^q$. In other words, if $\Delta < 0$, then the energy of the sequence $L^q$ is lower than the merit factor of the sequence $L$. Therefore, the merit factor of $L$ is going to be higher than the merit factor of $L^q$. More formally,
\begin{equation} 
\label{theorem1:property3}
\begin{split}
& \Delta < 0 \implies \mathbb{E}(L^q) - \mathbb{E}(L) < 0 \implies \mathbb{E}(L^q) < \mathbb{E}(L) \implies 2\mathbb{E}(L^q) < 2\mathbb{E}(L) \implies \\
& \implies \frac{1}{2\mathbb{E}(L^q)} > \frac{1}{2\mathbb{E}(L)} \implies \frac{n^2}{2\mathbb{E}(L^q)} > \frac{n^2}{2\mathbb{E}(L)} \implies \mathbb{MF}(L^q) > \mathbb{MF}(L). \\
\end{split}
\end{equation}

The derivative function allows us to reduce the memory complexity of some state-of-the-art algorithms from $O(n^2)$ to $O(n)$. In Table \ref{tab:memoryRequirements}, a comparison between the space required by the tau table and the memory requirement by the proposed method is presented. During the calculations, an assumption that both memory structures are comprised of integers (4 Bytes). For example, by using just one thread of the processors, the tau table corresponding to binary sequences with length $5000$ would require approximately $95.37$ Megabytes to be allocated for the tau table expansion routine, while the sidelobe array presented in this work would require the allocation of approximately $19.53$ Kilobytes. It should be emphasized, that interchanging the tau table used by the state-of-the-art algorithms with the proposed sidelobe array structure would not impact the time complexity of the tweaked algorithm. However, from a practical point of view, the significant memory reduction could greatly enhance the overall time performance of a tweaked algorithm, since the size of the sidelobe array could be usually saved inside the CPU cache layers, instead of saving it to the slower memory banks. Furthermore, interchanging the tau table with the proposed sidelobe array could allow the multithreading capabilities of modern CPUs, and even GPUs, to be fully utilized.

\begin{table}
\begin{center}
\caption{A comparison between the memory required by the tau table and the memory required by the proposed in-memory flip algorithm.}
\label{tab:memoryRequirements}
\ttfamily
\rowcolors{2}{light-gray}{light-cyan}
\begin{tabular}{lp{3cm}p{3cm}}
$n$ & {The memory required by using the tau table} & {The memory required by using the proposed method}   \\
\toprule
256 & 256.0 KB & 1.0 KB \\
512 & 1.0 MB & 2.0 KB \\
1024 & 4.0 MB & 4.0 KB \\
5000 &  95.37 MB & 19.53 KB \\
20000 &  1525.88 MB & 78.12 KB \\
99999 &  37.25 GB & 390.62 KB \\
\showrowcolors
\bottomrule
\end{tabular}
\end{center}
\end{table}

For example, we have implemented a lightweight version of the lssOrel algorithm \cite{bovskovic2017low} with the tau table reduced. The pseudo-code of the enhanced implementation is given in Algorithm \ref{algor:HasseOptimized}. The following notations were used:

\begin{itemize}
\item{$\Psi$ - a binary sequence with length $n$.}
\item{$\Omega_{\Psi}$ - the corresponding sidelobe array of $\Psi$ - the replacement of the tau table.}
\item{$\mathbb{H}$ - a set of fingerprints, or hashes, of visited candidates.}
\item{$\mathbb{T}_i$ - an inner threshold value. When the inner counter $w_i$ reaches $\mathbb{T}_i$, the set is flushed and the whole routine restarts. The threshold value $\mathbb{T}_i$ constrains the size of the set $\mathbb{H}$.}
\item{$\mathbb{T}_o$ - an outer threshold value. When the outer counter $w_o$ reaches $\mathbb{T}_o$, the program is terminated. However, $\mathbb{T}_o$ could be an expression as well. }
\item{$\mathbb{H}$.add(hash$(\Psi)$) - adding the hash of the binary sequence $\Psi$ to the set $\mathbb{H}$.}
\item{$C(\Omega_{\Psi})$ - the cost function, i.e. the sum of the squares of all elements in the sidelobe array $\Omega_{\Psi}$, which is equal to the energy of $\Psi$, or $\mathbb{E}(\Psi)$. }
\item{pickBestNeighbor$(\Psi, \Omega_{\Psi}, \mathbb{H})$ - a function, which returns the index of the best unexplored neighbor of $\Psi$, i.e. the binary sequence $\Psi^f$ with a distance of exactly 1 flip away from $\Psi$, s.t. hash$(\Psi^f)$ does not belong to the set $\mathbb{H}$. The pseudo-code of this helper function is given in Algorithm \ref{algor:pickBestNeighbor}.}
\end{itemize}

Several notations were used throughout the pseudo-code presentation shown in Algorithm \ref{algor:pickBestNeighbor}.

\begin{itemize}
\item{$\mathbb{MAX}$ - the maximum possible value, which the type of the variable \textbf{bestDelta} could hold. For example, if the variable \textbf{bestDelta} is of type \textbf{INT} (4 Bytes) then $\mathbb{MAX} = 7FFFFFFF_{16} = 2,147,483,647$}
\item{$\mathbb{P},\mathbb{Q}$ - two odd prime numbers, which are used to calculate the hash of the binary sequence. During our experiments, they were fixed to $\mathbb{P}=315223$ and $\mathbb{Q}=99041$. It should be noted, that no additional efforts were made to find better, in terms of hash collision false positives or false negatives rates, values of $\mathbb{P}$ and $\mathbb{Q}$.}
\end{itemize}

\algrenewcommand\algorithmicindent{0.5em}%
\begin{algorithm}[]
\caption{Lightweight flip probing of skew-symmetric binary sequences in linear both time and memory complexities}
\label{algor:InMemoryProbing}
\begin{algorithmic}[1]
\Function{Derivative}{$q, L, S$}
\State $\Delta = 0$
\State $\sigma = {(-1)}^{l-q}$
\For{$r =1; r < n-1; r += 2$}
		\If{$r \leq q$}
			\State \textbf{continue}
		\EndIf
		
		\State $\kappa = -8S_rL[q]$
		\State $\epsilon_1 = L[r-q-1]$
		\State $\epsilon_2 = L[q+r-n]$

		\If{$r < n-q$}
			\If{$q \neq r-q-1$}
				\State $\Delta = \Delta + 16 + \kappa\sigma\epsilon_1$	
			\EndIf		
		\Else
			\If{$q \neq r-q-1$}
				\State $\Delta = \Delta + 32 + \kappa(\epsilon_2 + \epsilon_1\sigma) + 32\epsilon_2\epsilon_1\sigma$		
			\Else 
				\State $\Delta = \Delta + 16 + \kappa\sigma\epsilon_2$
			\EndIf
		\EndIf
	\EndFor
\State \Return $\Delta$
\EndFunction
\end{algorithmic}
\end{algorithm} 

\algrenewcommand\algorithmicindent{0.5em}%
\begin{algorithm}[]
\caption{Heuristic algorithm, with tau table reduction, searching for binary skew-symmetric sequences with a high merit factor.  }
\label{algor:HasseOptimized}
\begin{algorithmic}[1]
\Procedure{MF}{$n, \mathbb{T}_i, \mathbb{T}_o$}

\State bestMF, $w_o \gets 0, 0$ 
\While {True}	 
	\State $\mathbb{H}, w_i, \gets \left\lbrace \emptyset \right\rbrace, 0$
	\State $\Psi \gets$ random  
	\State $\mathbb{H}$.add(hash$(\Psi)$)
	\State $V \gets C(\Omega_{\Psi})$   
	\While {True}	
		\State bestN $\gets$ pickBestNeighbor$(\Psi, \Omega_{\Psi}, \mathbb{H})$
		\If{bestN $== -1$}
			\State break
		\EndIf
		\State Flip(bestN, $\Psi, \Omega_{\Psi}$)
		\State $V \gets C(\Omega_{\Psi})$ 
		\State $w_i \pluseq 1$
		\State $\mathbb{H}$.add(hash$(\Psi)$)
		\If{$\frac{n^2}{2V} > $ bestMF}
			\State bestMF $\gets \frac{n^2}{2V}$
		\EndIf
		\If{$w_i > \mathbb{T}_i$}
			\State $w_o \pluseq 1$
			\State break
		\EndIf
	\EndWhile
\If{$w_o > \mathbb{T}_o$}
	\State break
\EndIf
\EndWhile
\EndProcedure
\end{algorithmic}
\end{algorithm}

\algrenewcommand\algorithmicindent{0.5em}%
\begin{algorithm}[]
\caption{Pseudo-code of the helper function \textbf{pickBestNeighbor}}
\label{algor:pickBestNeighbor}
\begin{algorithmic}[1]
\Function{pickBestNeighbor}{$\Psi, \Omega_{\Psi}, \mathbb{H}$}
\State bestN = -1
\State bestDelta = $\mathbb{MAX}$
\For{$q =0; q < \ceil{\frac{n}{2}}; q++$}
		\State $\delta$ = Derivative$(q, \Psi, \Omega_{\Psi})$
		\If{$\delta \leq$ bestDelta}
			\State hash=$\mathbb{P}$
			\For{$i =0; i < \ceil{\frac{n}{2}}; i++$}
				\If{$q == i$}
					\State hash=hash$*\mathbb{Q}$ - $\Psi[i]$
				\Else
					\State hash=hash$*\mathbb{Q}$ + $\Psi[i]$
				\EndIf
			\EndFor
			\If{hash$(\Psi) \in \mathbb{H}$}
				\State continue
			\EndIf		
			\State bestDelta = $\delta$
			\State bestN=$q$
		\EndIf

\EndFor
\State \Return bestN
\EndFunction
\end{algorithmic}
\end{algorithm}   

Algorithm \ref{algor:HasseOptimized} was implemented (C++) on a general-purpose computer equipped with a budget processor Xeon-2640 CPU, having a base frequency of 2.50 GHz. A skew-symmetric binary sequence with length 449 and a record-breaking merit factor of $6.5319$ was found after approximately one day. It should be noted that all 12 threads of the CPU were launched in parallel. As a comparison, the currently known optimal results (a merit factor of 6.5218) were acquired by using the Slovenian Initiative for National Grid (SLING) infrastructure (100 processors) and 4-day threshold limitation \cite{bovskovic2016github}. The binary sequence is given in a hexadecimal format in Table \ref{tab:meritFactor449}.

It should be emphasized that the flip operation for the middle index of the skew-symmetric binary sequence $\Psi$ is not permitted. However, this is not affecting the search space by cutting some parts of it. Indeed, let's define the binary sequence $\mathbb{B} = b_1b_2 \cdots b_{l}Mb_{l+1}b_{l+2} \cdots b_{2l}$ of length $n=2l+1$ and the binary sequence $\overline{\mathbb{B}}$ as the binary sequence $\mathbb{B}$ with all the bits flipped, i.e. $\overline{\mathbb{B}} = \overline{b_1}\overline{b_2} \cdots \overline{b_{l}}\overline{M}\overline{b_{l+1}}\overline{b_{l+2}} \cdots \overline{b_{2l}}$. It could be easily shown that all sidelobes of  
$\mathbb{B}$ and $\overline{\mathbb{B}}$ are identical. Indeed, $$C_u(\overline{\mathbb{B}})=\sum_{j=0}^{n-u-1} \overline{b_j}\overline{b_{j+u}} = \sum_{j=0}^{n-u-1} (-1)^2b_jb_{j+u} = C_u(\mathbb{B}).$$ 
\begin{table}
\begin{center}
\caption{An example of a skew-symmetric binary sequence with length 449 and a record merit factor found by Algorithm \ref{algor:HasseOptimized}. The sequence is presented in HEX with leading zeroes omitted.}
\label{tab:meritFactor449}
\ttfamily
\rowcolors{2}{light-gray}{light-cyan}
\begin{tabular}{lp{5cm}ll}
$n$ & Sequence in HEX & MF   \\
\toprule
449 & 
96f633d86fe825794ed23a9dfd7d4c3
abd080cf76cbf9bdab9a7b2533e3161
901d1950c774ca8bd012cfd7d5d8123
c4f97e285469d327478 & 6.5319 \\

\showrowcolors
\bottomrule
\end{tabular}
\end{center}
\end{table}

\section{On the Bernasconi Conjecture}
\label{sec:fitness}

As discussed throughout the introduction section, in \cite{bernasconi1987low} Bernasconi conjectured that stochastic search procedures will not yield merit factors higher than 5 for long sequences (greater than 200). It should be mentioned that this prediction was made in 1987. Since then, many years have passed and pieces of evidence that stochastic search procedures could perform better than the prediction's expectations were found. Indeed, heuristic algorithms that could found odd binary sequences with lengths up to about 500 and merit factors greater than 5 were discovered. However, the Bernasconi conjecture appears valid when the threshold of the binary sequence's length is updated and lifted. Since during the last 35 years the computational capabilities of modern CPUs are rising almost exponentially such actualization would be fair. However, if a stochastic search procedure is found, a procedure that could reach extremely long binary sequences with merit factors greater than 5, by using a mid-range general-purpose computer, then the barriers predicted by Bernasconi could be very pessimistic.

Some more experiments were made by using Algorithm \ref{algor:HasseOptimized} and skew-symmetric binary sequences with lengths greater than 1000. For example, within several seconds, a binary sequence with length 1001 and a merit factor greater than 5 was discovered. By leaving the routine for a minute, binary sequences with merit factors up to 5.65 were reached. Then, within several seconds as well, a binary sequence with length 2001 and merit factor greater than 5 was discovered. However, this time the routine needed almost an hour to reach binary sequences with merit factors up to 5.40. When the length is increased to 5001, the algorithm required half a day to reach a binary sequence with MF greater than 5.10. Finally, the algorithm failed to reach a binary sequence with length 10001 and a merit factor greater than 5 within 24 hours (by using all the twelve threads of the processor). The numerical experiments suggest that Algorithm \ref{algor:HasseOptimized} is not able to find binary sequences with lengths greater than 10000 and merit factor greater than 5. 

Indeed, the Algorithm \ref{algor:HasseOptimized} property of avoiding Hasse cycles, or the self-avoiding walk (SAW) property, yields binary sequences with near-optimal merit factors. However, the efficiency of this strategy melts away when binary sequences with bigger lengths are used. This is not surprising, since the bigger the length, the larger the search space is. For example, the search space of the set of all skew-symmetric binary sequences with length 10001 is $2^{5001}$. More importantly, several more computational burdens were introduced by Algorithm \ref{algor:HasseOptimized} itself:

\begin{itemize}
\item{The \textbf{pickBestNeighbor} function (see Algorithm \ref{algor:pickBestNeighbor}) is looking for the best neighbor of the current binary sequence $\Psi$. Thus, each calling of the function would trigger the \textbf{Derivative} function exactly $n$ times.}
\item{As previously discussed, Algorithm \ref{algor:HasseOptimized} is using a hashing technique to keep an unordered set of the already visited notes. Such approach is causing a significant computation burden to the algorithm for larger values of $n$:

\begin{enumerate}
\item{The unordered set strategy requires at least $\mathbb{G}\mathbb{X}n\mathbb{T}_o$ bytes of memory, where $\mathbb{G}$ is the count of the threads used by the processor, while $\mathbb{X}$ is the size in bytes of the used variable type.}
\item{Frequently, when a candidate $\Psi^q$ with lower score $\delta$ is found (see line 6 from Algorithm \ref{algor:pickBestNeighbor}), a hash of the candidate should be calculated, so to be further checked was the binary sequence $\Psi^q$ met before.}
\end{enumerate} }
\end{itemize}

To annihilate all the aforementioned computational burdens, an Algorithm \ref{algor:TotalReduction} is proposed. In summary, the following simplifications were introduced:

\begin{enumerate}
\item{The \textbf{pickBestNeighbor} function straightforwardly accept the first met neighbor having a strictly better score.}
\item{By introducing the previous tweak, the algorithm cycle trapping is avoided. It should be noted that if small values of $n$ are used, this could greatly worsen the quality, in terms of the high merit factor, of the binary sequences found. However, when considering larger values of $n$, the numerical experiments suggest that this tweak could be highly efficient. Thus, the need of using unordered set could be completely annihilated and the memory complexity of the algorithm significantly reduced.}
\item{Since the unordered set was annihilated, the hash routines are removed as well.}

\end{enumerate}

\algrenewcommand\algorithmicindent{0.5em}%
\begin{algorithm}[]
\caption{A heuristic algorithm, with tau table, unordered set, and hashing routines reduced, for searching long skew-symmetric binary sequences with a high merit factor. Both the time and memory complexity of the algorithm are  $O(n)$.}
\label{algor:TotalReduction}
\begin{algorithmic}[1]
\Procedure{SHC}{$n, \mathbb{T}$}

\State $\Psi \gets$ random 
\State {$V^*$, $V$, $\mathbb{G}$, $\mathbb{L}$, $c$ $\gets $ $C(\Omega_{\Psi})$, 0, True, False, 0}
\While {$c < \mathbb{T}$}
	\State $c \pluseq 1$
	\If{$\mathbb{G}$}
		\State pick random $r \in \left[0, \floor{\frac{n}{2}}\right)$
		\For{$q \in \left[0, \floor{\frac{n}{2}}\right)$}
			\State $\delta$ = Derivative$((r+q) \bmod \floor{\frac{n}{2}}, \Psi, \Omega_{\Psi})$	
			\If{$\delta > 0$}
				\State \textbf{continue}
			\EndIf		
			\State Flip($(r+q) \bmod \floor{\frac{n}{2}}, \Psi, \Omega_{\Psi}$)
			\State $V \pluseq \delta$
			\If{$V^* >  C(\Omega_{\Psi})$}
				\State $V^*$, $\mathbb{L}$ $\gets C(\Omega_{\Psi})$, True				
				\State \textbf{break}
			\Else
				\State Flip($(r+q) \bmod \floor{\frac{n}{2}}, \Psi, \Omega_{\Psi}$)
			\EndIf
		\EndFor
		\If {$\mathbb{L}$}
			\State $\mathbb{G}$, $\mathbb{L}$ $\gets$ True, False
			\State \textbf{continue}
		\Else
			\State $\mathbb{G}$ $\gets$ False
		\EndIf
	\Else
		\State Quake($\mathbb{Q}, \Psi, \Omega_{\Psi}$)
		\State $\mathbb{G}$, $\mathbb{L}$ $\gets$ True, False
	\EndIf
\EndWhile
\EndProcedure
\end{algorithmic}
\end{algorithm} 
 
In Algorithm \ref{algor:TotalReduction} the following notations were used:
\begin{itemize}
\item{$\mathbb{T}$ - the threshold value of the instance.}
\item{$C$ - the cost function.}
\item{$V$, $V^*$ - respectively the current best and the overall best score values.}
\item{$c$ - the counter. The algorithm quits if the counter $c$ reaches the threshold $\mathbb{T}$.}
\item{$\mathbb{L}$, $\mathbb{G}$ - binary variables: $\mathbb{L}$ (local) is activated if $V$ is improved, while $\mathbb{G}$ (global) is activated if $V^*$ is improved.}
\item{Quake function - the function flips $\mathbb{Q}$ random bits in $\Psi$.}
\end{itemize}

During our experiments, by using Algorithm \ref{algor:TotalReduction}, we were able to reach skew-symmetric binary sequences with lengths up to 100,001 and merit factors greater than 5. However, the greater the length of the binary sequence is, the larger the value of $\mathbb{Q}$ should be. Some of those $\mathbb{Q}$ values, used during our experiments, are given in Table \ref{tab:quakesMeasurements}. It should be emphasized, that those $\mathbb{Q}$ values guarantee to reach a skew-symmetric binary sequence with merit factors greater than 5.0, but it is highly unlikely that exactly those values would yield the best results. 

For example, by using Algorithm \ref{algor:TotalReduction}, a binary sequence with length 10,001 and merit factor greater than 5 was reached for approximately one minute. Leaving the algorithm for another minute would reach merit factors of 5.10 and higher. Doubling the length of the binary sequence to 20,001 required from Algorithm \ref{algor:TotalReduction} approximately 4 minutes to reach a skew-symmetric binary sequence with a merit factor greater than 5.

Binary sequences with length 50,001 and a merit factor greater than 5 were reached for leaving the algorithm for approximately 40 minutes, while binary sequences with length 100,001 and a merit factor greater than 5 were reached for approximately 5 hours. However, it should be emphasized that the larger the sequence, the larger the number of quakes $\mathbb{Q}$ should be. In Table \ref{tab:quakesMeasurements} the values of $\mathbb{Q}$ corresponding to the binary sequences' lengths used throughout the experiments are given. Small cuts from the history of the search traces are provided within the four complimentary files. Each file holds skew-symmetric binary sequences of fixed length - ${2}^4{5}^4+1$, ${2}^5{5}^4+1$, ${2}^4{5}^5+1$ or ${2}^5{5}^5+1$. All sequences posses merit factors greater than 5.
\begin{table}[h]
\begin{center}
\caption{The number of quakes used throughout our experiments.}
\label{tab:quakesMeasurements}
\ttfamily
\rowcolors{2}{light-gray}{light-cyan}
\begin{tabular}{r|l}
Length $n$ & Quake $\mathbb{Q}$  \\
\toprule
999 & 1  \\
1499 & 2  \\
1999 & 3  \\
2999 & 4 \\
4999 & 6  \\
10001 & 14  \\
20001 & 30  \\
50001 & 70  \\
100001 & 160  \\
\showrowcolors
\bottomrule
\end{tabular}
\end{center}
\end{table}

\begin{figure}[h]
    \centering
    \includegraphics[width=0.5\textwidth]{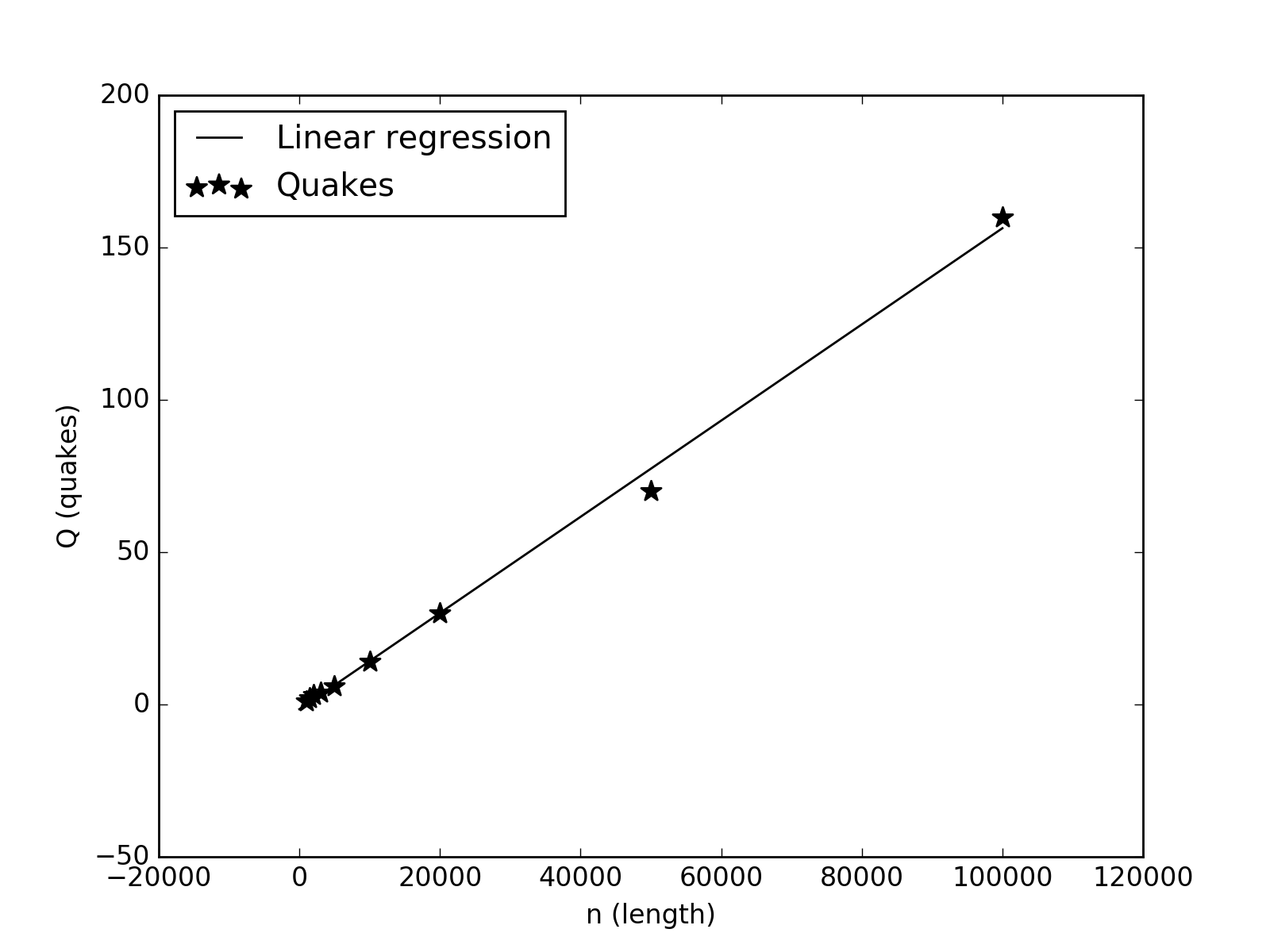}
    \caption{A linear regression made to all the $(n,\mathbb{Q})$ pairs from Table \ref{tab:quakesMeasurements}. The equation representing the linear fit is $\mathbb{Q}=0.001578787n - 1.546093$.}
    \label{fig:LinearFitQ}
\end{figure}

\begin{figure}[h]
    \centering
    \includegraphics[width=0.5\textwidth]{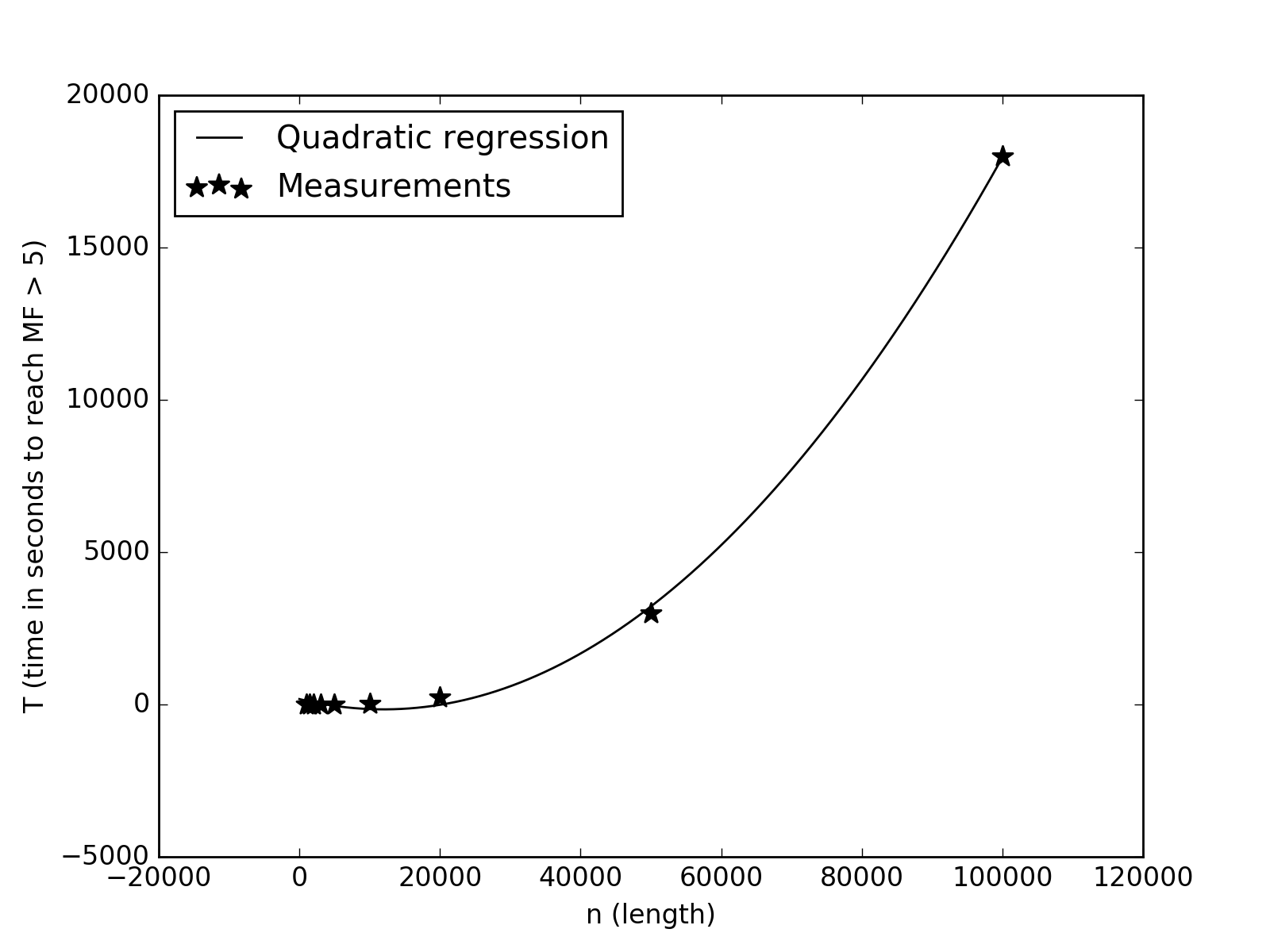}
    \caption{A quadratic regression made to all the $(n,\mathbb{T})$ measurements. The equation representing the quadratic fit is $\mathbb{T}=177.2867 - 0.0562043n + 0.000002340029n^2$.}
    \label{fig:QuadraticFitT}
\end{figure}

The numerical experiments suggest that value of $\mathbb{Q}$ grows linear with the length of the binary sequence. This is clearly visible in Figure \ref{fig:LinearFitQ}. The time required (in seconds) to reach binary sequences with a merit factor strictly greater than 5 are given in Figure \ref{fig:QuadraticFitT}. As expected, the time required to reach a binary sequence with merit factor greater than 5 grows quadratic with the size of the binary sequences $n$. 

Both the regression models are rough approximations of the algorithm's behavior. For a more precise estimation - more instances of the algorithm should be analyzed. However, one very important property of Algorithm \ref{algor:TotalReduction} should be further highlighted. When a counter to the function \textbf{Quake} is attached, during the optimization routine a total of approximately 2000-2500 calls to the function are made  before a binary sequence with merit factor greater than 5 is reached. This observation, as well as the numerical pieces of evidence found through our experiments, suggest that given a arbitrary binary sequence $\mathbb{B}$ with length $n$, and by using general-purpose computer with 12 threads, as well as C++ implementation of Algorithm \ref{algor:TotalReduction} launched with variable $\mathbb{Q}$ close to $\ceil{0.001578787n - 1.546093}$, $\mathbb{B}$ could be optimized to a binary sequence with merit factor greater than $5$, after an approximately $177.2867 - 0.0562043n + 0.000002340029n^2$ seconds.

\section{Conclusions}

In this work, by using some mathematical insights, an alternative to the tau $\Tau(S)$ table, which was frequently utilized by state-of-the-art algorithms designed to search skew-symmetric binary sequence with high merit factor, is suggested. The proposed algorithm could be used to reduce the memory complexity of the skew-symmetric binary sequences' flip operation from $O(n^2)$ to $O(n)$. Thus, the technical limitations of stochastic algorithms searching for skew-symmetric binary sequences with high merit factors are now significantly reduced. Finally, a heuristic method for constructing skew-symmetric sequences of arbitrary length and merit factor greater than 5 is proposed. Numerical experiments are provided for some chosen lengths up to $10^5+1$.




\bibliographystyle{IEEEtran}
\bibliography{refs}

\end{document}